# AC susceptibility of $YBa_2Cu_3O_{7-\delta}$ superconducting thin film in a perpendicular field

H. Salamati*, P. Kameli
Solid State Lab. Dept. of Physics
Isfahan University of Technology, Isfahan, Iran

---

## Abstract

The temperature dependence of ac susceptibility of $YBa_2Cu_3O_{7-\delta}$ thin film was measured as a function of frequency and ac field amplitude in a field perpendicular to the film plane. We show that all measured ac susceptibility curves can be scaled into a single curve using the peak temperature ($T_p$) of its imaginary part as the scaling parameter. The frequency effect on ac susceptibility was also measured. As the frequency increases, ($T_p$) shifts to higher temperature. This effect can be interpreted in terms of flux creep. The temperature T, and field, $H_{ac}$, dependence of the activation energy obtained from the Arrhenius plot for the frequency ($f$) and ($T_p$), can be described as $U(H_{ac},T) = U_0(1-T/T_c)H_{ac}^{-0.18}$.

Keywords: thin film, Ac Susceptibility, Scaling, Activation energy

---

*Corresponding author.
E-mail address: salamati@cc.iut.ac.ir
Fax: 98-311-3912376

## 1. Introduction

In addition to electrical resistivity, ac susceptibility ($\chi' + i\chi''$) is widely used as a nondestructive method for the determination and characterization of the intergrain component in high-$T_C$ thin film superconductors. Experimentally, the real part of susceptibility, $\chi'$, is observed to be stepwise as a function of temperature, representing transition from near-perfect screening to complete penetration of external ac magnetic field into the sample while the peak observed in the imaginary part of susceptibility, $\chi''$, is a measure of dissipation in the sample [1]. Ac losses in type two superconductors arise from viscous flux flow motion and hysteresis due to pinning of flux vortices [2]. Thus, pinning losses are independent of frequency but dependent on the field, whereas flux flow losses depend on the frequency not on the field. Experimentally, both field and frequency dependencies can be observed simultaneously [2-5]. Many researchers have observed that the complex ac susceptibility of high-$T_c$ superconductors is weakly dependent on the frequency [6-8] (relative to field dependence) and strongly on the ac magnetic field amplitude $H_{ac}$ [9-11].

It has been shown that the dependence of ac susceptibility on the ac field amplitude, $H_{ac}$, was determined by the pinning strength parameter [11,12]. Critical state models have been very successful in accounting for the major features of the curves of $\chi_{ac}$ vs. T and their evolution with $H_{ac}$ [9-13].

In this paper, we report on the measurements of ac susceptibility of $YBa_2Cu_3O_{7-\delta}$ (YBCO) thin film for various ac field $H_{ac}$ and frequencies. We found that all $\chi'(T)$ and $\chi''(T)$ curves for various $H_{ac}$ can be scaled in to a single curve using the $\chi''$ peak temperature $T_p$ as a scaling parameter. We also calculate the activation energy U ($H_{ac}$, T) at grain boundaries from the Arrhenius plot for the frequency, f, and the peak temperature of imaginary part of susceptibility, $T_p$.

## 2. Experimental

A high quality c-axis-oriented YBCO epitaxial thin film was prepared by Pulsed Laser Deposition of stoichiometric YBCO target on a $SrTiO_3$(100) substrate. The X-Ray diffraction showed no other observable peaks than (00l) and the full width at half maximum of the rocking curve for the (005) peak was less than $0.3^{\circ}$, indicating a good epitaxial c-axis orientation. The ac susceptibility measurements were made using Lakeshore ac susceptometer Model 7000, with the field perpendicular to the film plane. The temperature dependence of ac susceptibility was measured for various ac fields $H_{ac}$ (given as rms values) from 0.5 to 10 Oe and frequencies f, from 10 to 1000Hz. For all measurements, the sample was cooled down in zero fields and then warmed up at a constant rate of 0.2 k/min.

## 3. Results and discussion

### 3.1. Scaling of the ac susceptibility

Fig.1 shows the temperature dependencies of the normalized real and imaginary parts of ac susceptibility for various ac fields at a frequency of 1kHz. . The diamagnetic onset temperature of intrinsic superconducting transition is about 90k. It is clear from Fig.1 that, as the field amplitude $H_{ac}$ increases, the peak of $\chi''$(T) shifts to lower temperatures and broadens. The degree of shift at various $H_{ac}$ is proportional to the magnitude or strength of the pinning force. The weaker the pinning, the smaller the critical current.

Fig.2 shows the variation of peak temperature, $T_p$ as a function of ac field amplitudes, $H_{ac}$. Although $H_{ac}$ was plotted vs. $T_p$, it can also be related as a $J_c$ vs. temperature curve. Brandt [15], using the Critical State Model, showed that for thin film with strip geometry in perpendicular field, the peak of $\chi''$ takes place when the ac field amplitude reaches:

$$H_p = J_c \frac{d}{\pi}\left[1 + \ln\frac{2a}{d}\right] \quad (1)$$

where d is the thickness and 2a is the width of thin film. Using above equation, Fig.2 can be considered as a graph of $J_c$ vs. temperature. Our data is in good agreement with $J_c$ dependence of temperature as $J_c \sim (1 - T/T_c)^{2.23}$.

Herzog et al [15] obtained the following expressions for the real and imaginary parts of ac susceptibility for a strip shape sample:

$$\chi' = -\frac{1}{h}\tanh h \quad (2a)$$

$$\chi'' = -\frac{1}{h}\tanh h + \frac{2}{h}\tanh\frac{h}{2} \quad (2b)$$

with $h = \frac{kH_{ac}}{J_c d}$, where k is a constant. These equations show that we can scale the $\chi'$(T) and $\chi''$(T) curves at various ac fields, $H_{ac}$, by using h as an independent variable. Han et al [5] found that all plots of $\chi'$ and $\chi''$ vs. temperature, for Bi2223/Ag tape in perpendicular ac field, can be scaled by the scaling feature using $T_p$ as the scaling parameter. However, in their $\chi''$(T) curves, the scaling shows deviation at low temperatures. As seen in Fig.3, in our curves this deviation is not present. This deviation may be related to the eddy current component induced in the silver sheath. Savvides et al [16] showed that at higher frequencies (f >50Hz) and low temperatures, the eddy current loss of silver sheath in Bi2223/Ag tape becomes predominant. In other words, the critical state model cannot be used to account for the ac susceptibility behavior in such a case.

### 3.2. Temperature and field dependence of U ( $H_{ac}$ , T )

High-$T_c$ superconductors response to an external ac field can be described by one of the following three different regimes [3]: (1) a critical-state regime with no detectable frequency dependence but large ac field dependence, (2) a linear ohmic regime with eddy current frequency dependence but no ac field dependence and (3) an intermediate flux-creep regime with both frequency and ac field dependent ac susceptibility. The first two regimes are relatively well understood and can, in principle, be regarded as limiting cases of the third more complex intermediate regime. By varying the frequency, the effect of flux creep on the ac susceptibility, can be investigated. Fig.4 shows the temperature dependencies of $\chi'$ and $\chi''$ for $H_{ac}$ =5Oe at frequencies f =111,285,666 and 1000Hz.The effect of frequency can be clearly observed. As frequency increases, the susceptibility curves (in particular peak in $\chi''$) shift to higher temperature. As explained by Müller [8], when the frequency is increased, the intergranular vortices have less time (relative to low frequency) to relax and then penetrate the superconductor during each cycle, according to the flux creep model. In order to reach full flux penetration, the effective intergranular pinning force density must be weakened. Since the pinning force density weakens with increasing temperature, the peak temperature, $T_p$, must increase with increasing frequency. From the shift of $T_p$ with frequency, one can derive U ( $H_{ac}$ ,T).

The maximum of $\chi''$(T) is at a frequency given by the following Eq. [6,14]:

$$f = f_0 \exp(-U/k_B T) \qquad (3)$$

where $f_0$ is an attempt frequency and U is the activation energy for flux creep.

Fig. 5 shows exponential dependence of f vs. $1/T_p$ in our experimental range. From this Arrhenius-like equation, activation energy can be calculated for each field amplitude $H_{ac}$. Fig. 6 shows U as a function of magnetic field $H_{ac}$. From Fig.6, we found a power law dependence of $U(H_{ac}) \propto H_{ac}^{-0.18}$. It should be emphasized that the Arrhenius curve slopes are directly related to, but not equal to, the activation energies [8]. Namely, using a typical slope of 5ev at $H_{ac}$ =4Oe and $T_p$ =87k as activation energy gives a $f_0$ =f exp ($U/k_B$ $T_p$)=f exp (5/0.0076) $\sim 10^{280}$, if f is in the order of the measure frequency. Clearly, the true activation energy must be much smaller in order to obtain a $f_0$ value, which has physical meaning. But it is known that the activation energy is temperature dependent and the activation energy U extracted from plot of $1/T_p$ vs. ln f is thought to be equal to the activation energy at T=0, U ($H_{ac}$, T=0). Considering its temperature dependence, the activation energy is rewritten as the phenomenological relation: U ($H_{ac}$, T) = U ($H_{ac}$, 0)(1-T/$T_c$), which we can also obtain a linear $1/T_p$ vs. ln f plot. Of course, the slope would change to -$k_B$/U ($H_{ac}$, 0) and the constant term to $k_B$ /U ($H_{ac}$, 0) ln $f_0$ +1/$T_c$. The value of $f_0$ determined in this case is ($\sim 10^8$), which is of the order of magnitude expected for the hopping frequency between potential wells ($10^7$ - $10^{12}$) [8]. From both U ($H_{ac}$) and U (T) dependences mentioned above, it might be concluded that the effective activation energy in these ac field and frequency ranges could be presented with the formula: U ($H_{ac}$, T)= $U_0$ (1-T/$T_c$) $H_{ac}^{-0.18}$ for $YBa_2Cu_3O_{7-\delta}$ thin films.

## 4. Conclusion

Frequency and field dependencies of the ac susceptibility of YBCO thin film in the perpendicular ac field have been investigated. The real ($\chi'$) and imaginary ($\chi''$) parts of complex ac susceptibility ($\chi' + i\chi''$) show a scaling behaviour for various ac field amplitudes. The peak temperature $T_p$ of the imaginary part ($\chi''$) shifts slightly to higher temperature with increasing frequency. The frequency dependence of ac susceptibility can be attributed to intergranular flux creep or magnetic relaxation. The activation energy obtained from an Arrhenius plot for the frequency f and peak temperature $T_p$ in the experimental ranges could be described by the formula $U(H_{ac}, T) = U_0 (1\text{-}T/T_c) H_{ac}^{-0.18}$ for temperature T, and ac field, $H_{ac}$.

## Acknowledgments

The authors would like to thank Isfahan University of Technology for supporting this project.

## Figure captions

Fig.1. Temperature dependence of normalized real (a) and imaginary (b) parts of ac susceptibility for various ac field amplitudes at a frequency of 1kHz.

Fig.2. Field dependence of peak temperature ($T_p$).

Fig. 3. Scaling behavior of real (a) and imaginary (b) parts of ac susceptibility for curves represented in Fig.1.

Fig. 4. Temperature dependence of real (a) and imaginary (b) parts of ac susceptibility for four selected frequencies

Fig 5. Plot of $1/T_p$ vs. f . The activation energy U, is given by the slope.

Fig. 6. Ac magnetic field dependence of activation energy (U).

Fig.1
H. Salamati et al

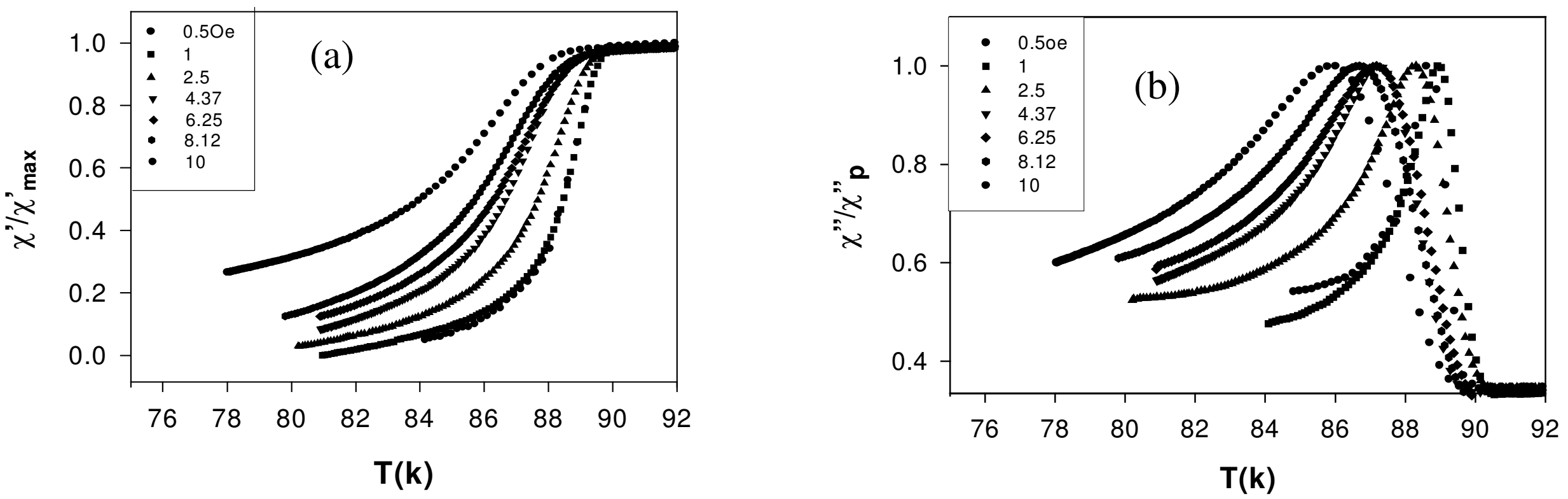

Fig.2
H. Salamati et al

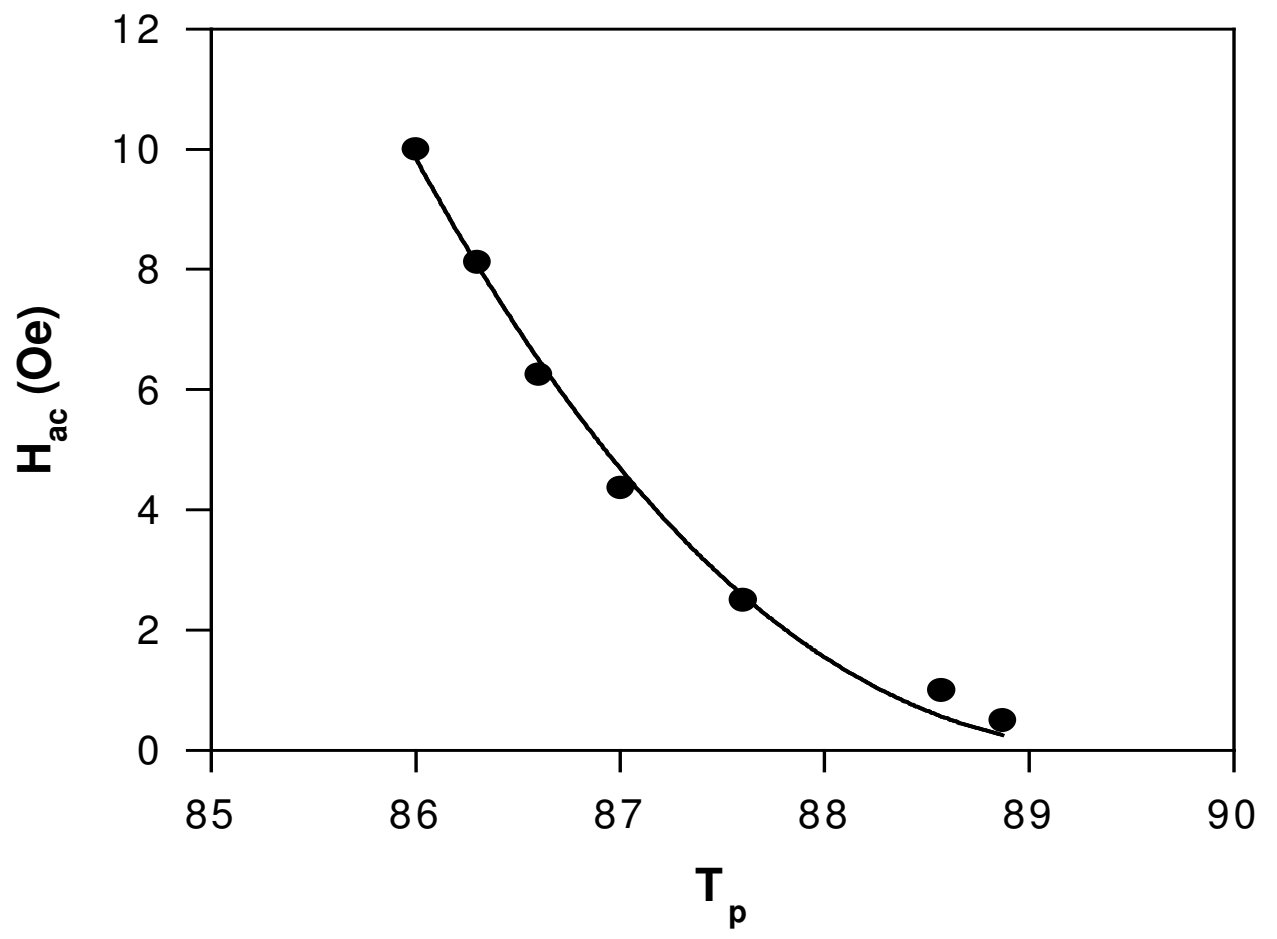

Fig.3
H. Salamati et al

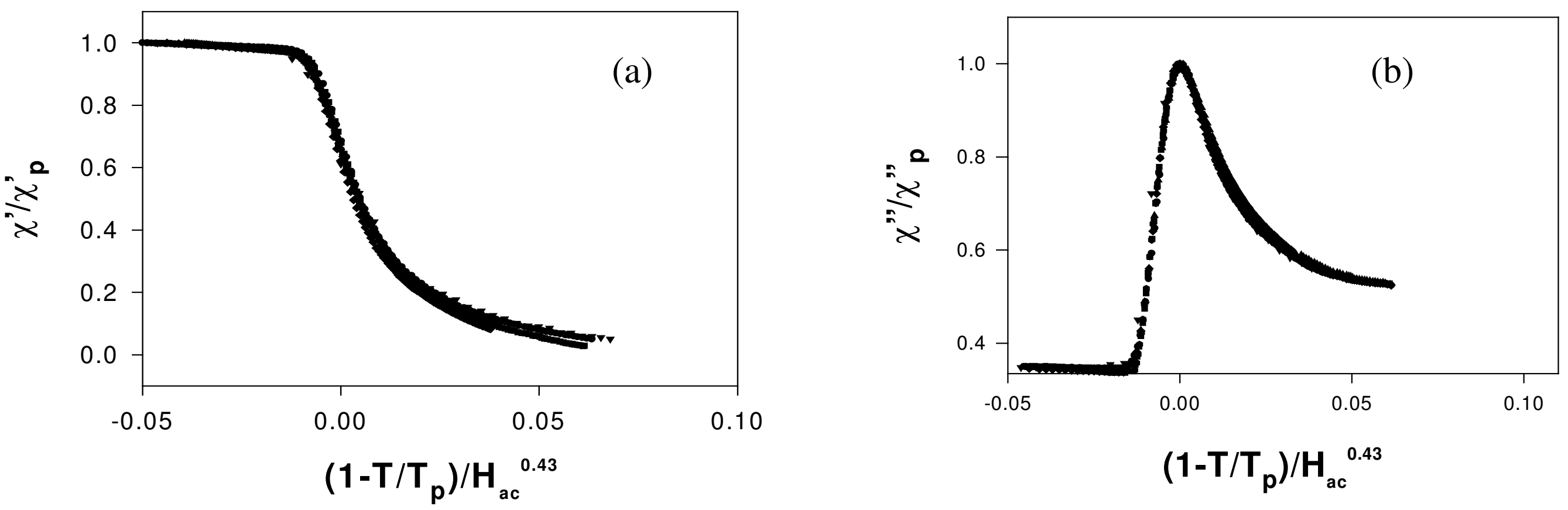

Fig.4
H. Salamati et al

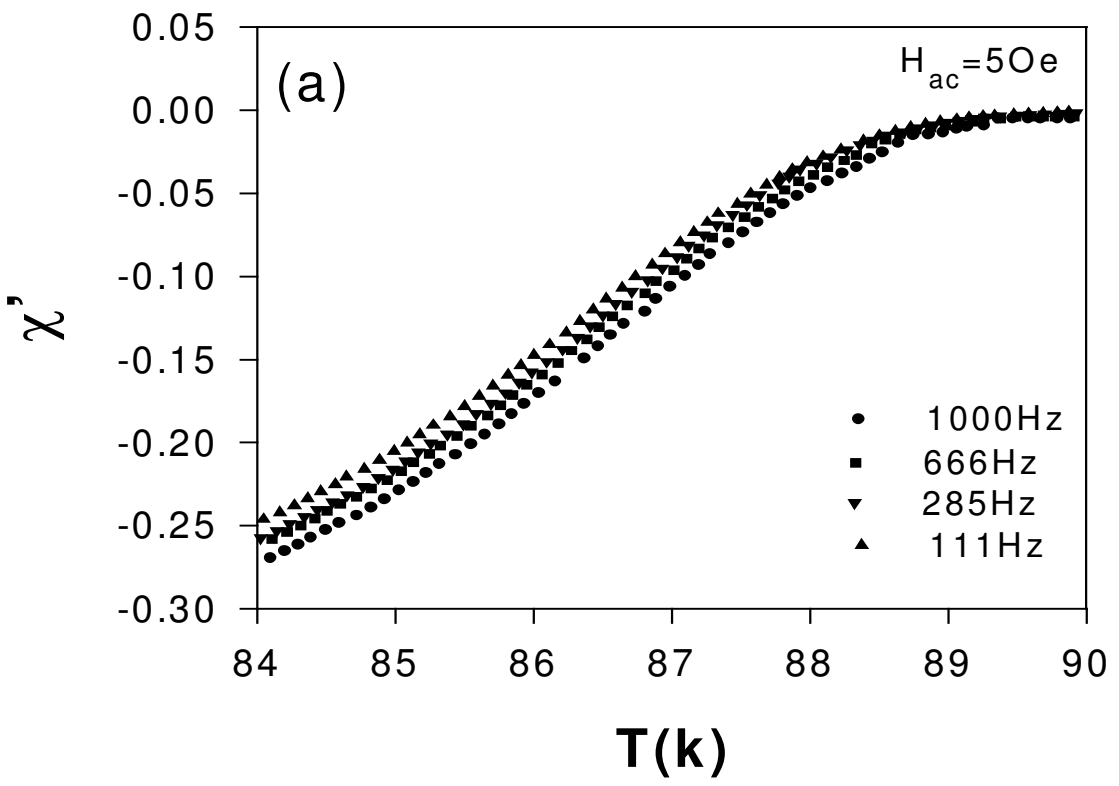

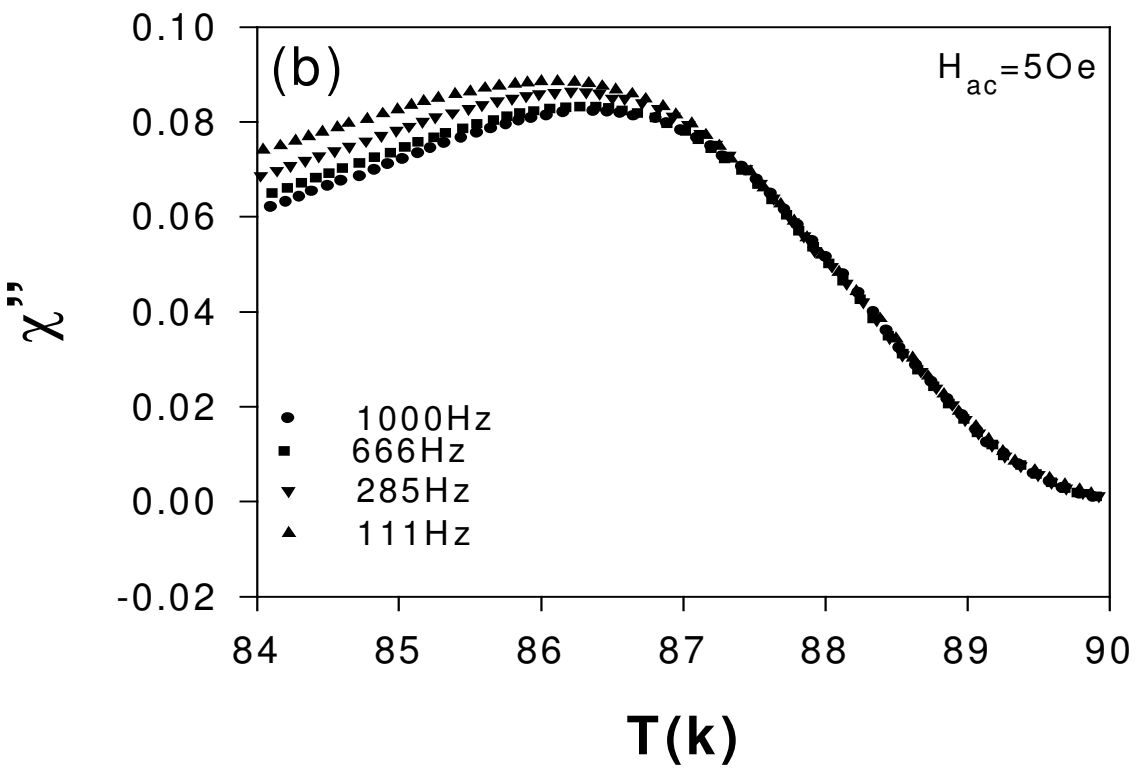

Fig.5
H. Salamati et al

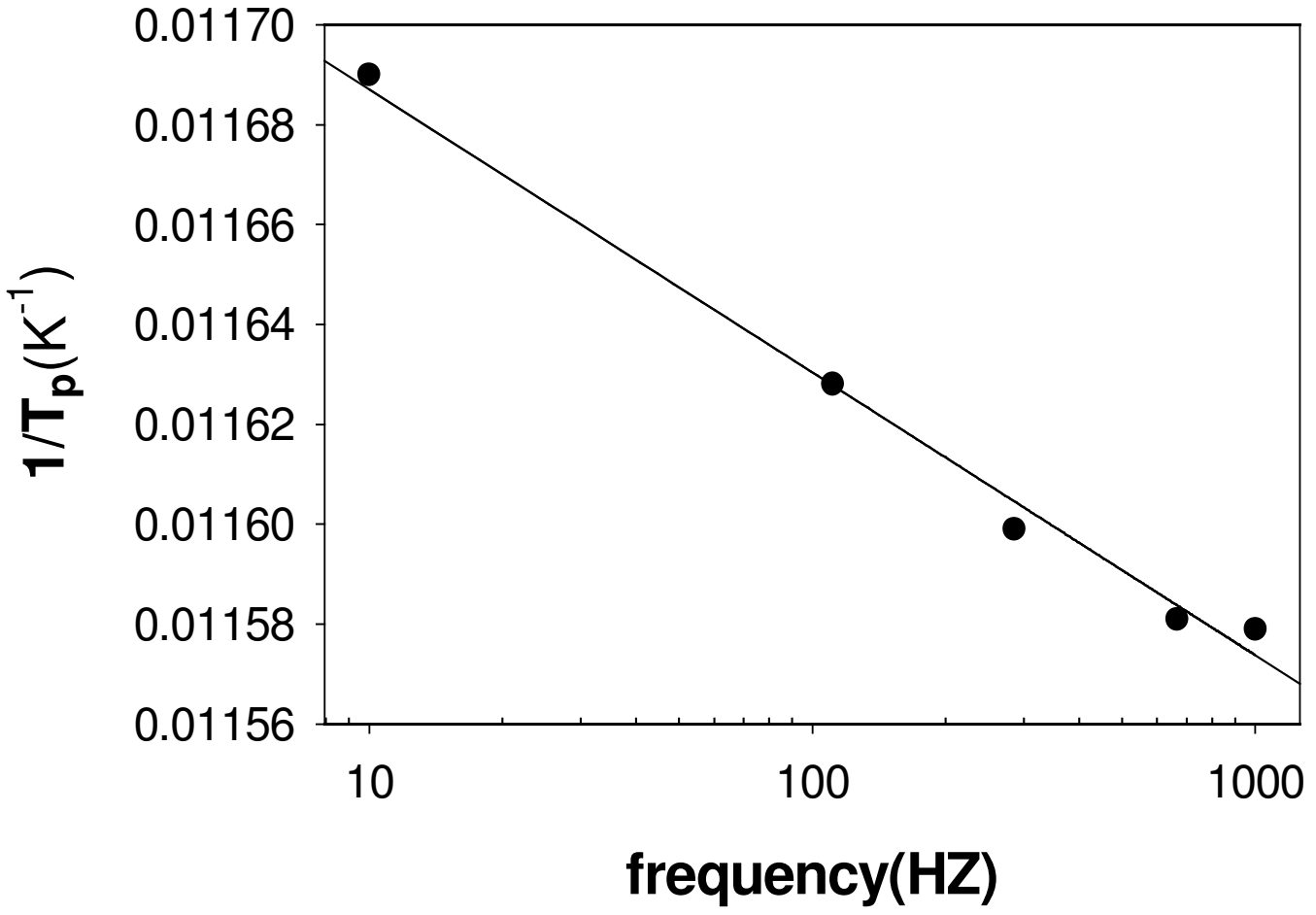

Fig.6

H. Salamati et al

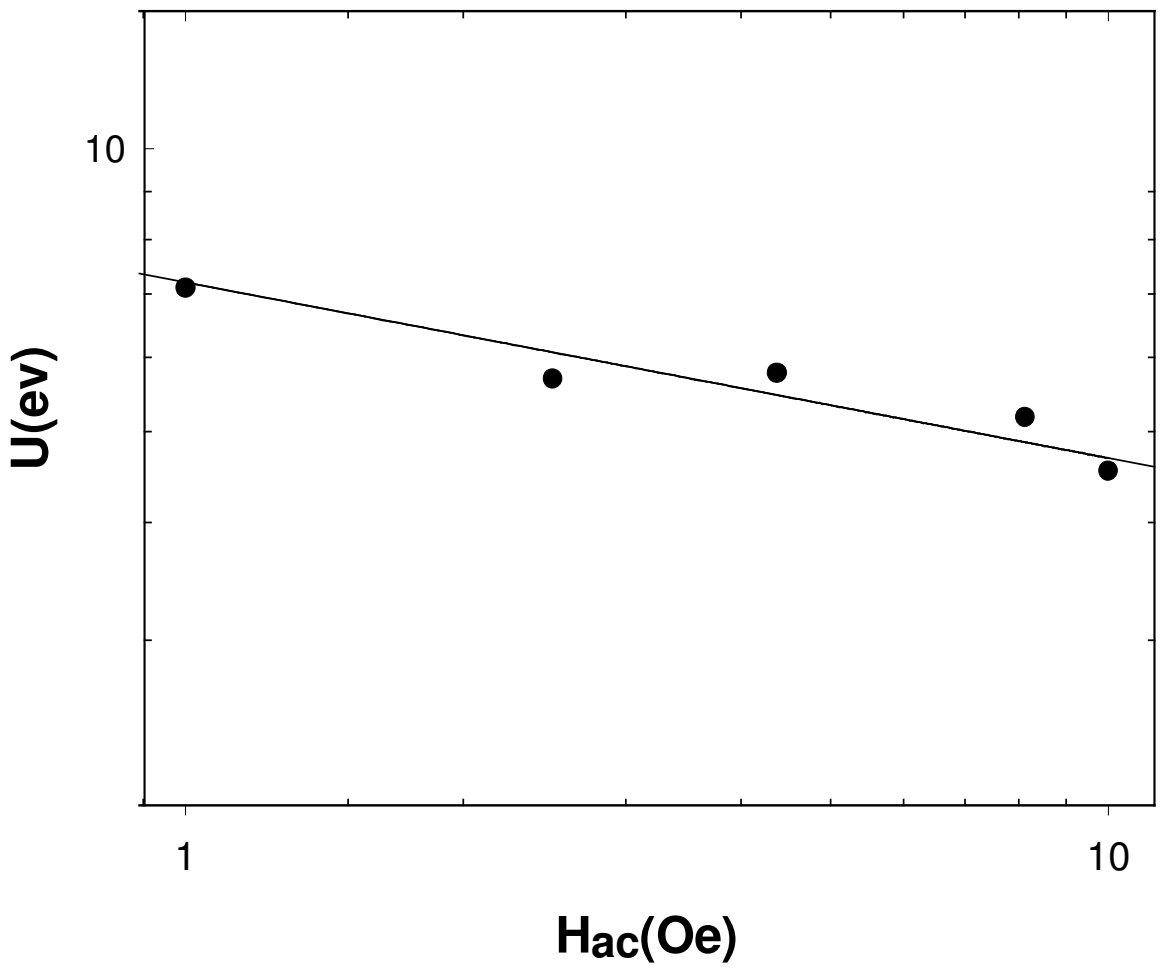